\documentclass[twocolumn,showpacs,preprintnumbers,amsmath,amssymb]{revtex4}

\usepackage{graphicx}
\usepackage{dcolumn}
\usepackage{bm}

\begin{document}
\title{Statistical description of Universe formation.}
\author{B. I. Lev and A.G. Zagorodny}

\affiliation{Bogolyubov Institute for Theoretical Physics, NAS,
Ukraine, Meytrologichna 14-b, Kyiv 03143, Ukraine.}
\date{\today}
\pacs{73.21.Fg, 78.67.De}

\begin{abstract}
Statistical description of Universe as non - equilibrium system has been proposed. Based on two fundamental principles: 
law increasing the entropy and minimization of energy of system is defined condition the Big Bang and arose the Universe. 
\pacs { 05.70.Rr,05.20.Dd,05.40.-a}
\end{abstract}
\maketitle

\section{Introduction}
 Within the context of the fundamental principles of thermodynamics, any macroscopic system embedded in a thermal 
 bath approaches equilibrium  during some relaxation time. Inthe equilibrium state, the properties of the system 
 do not depend on the way of how the equilibrium has been established. The equilibrium state is, however, realized 
 only under certain idealized conditions, so in  reality the properties of the system in a quasi-stationary (steady) 
 state can  depend both on the specifics of the  interaction  of the system with the thermal bath and the characteristics 
 of the bath \cite{Huang}- \cite{Lan}. The same concerns nonequilibrium systems. But our Universe are non-equilibrium 
 \textsl{a priory}. The main idea this letter are in
possible approach to answer on question from what state and for inner conditions of vacuum can be arose of Universe 
and what can be are motive the Big Bang. We not pretended on fully solution this problem but would like appeal the 
attention on one not difficult way take into account the non-equilibrium condition existence Universe. Currently, 
there does not exist a well-developed description method of the non-equilibrium system but although far from equilibrium 
systems are abundant in nature, there is no unified commonly accepted theoretical approach which determines possible 
states of such systems. Hence, it is a fundamentally important task to develop a method for exploring general properties 
of stationary states of open systems and to establish conditions of their existence. The way solution such general problem 
was presented already in Gibbs approach \cite{Gibbs}. It is first step to solution the previously formulate problem. 

\section{Statistical approach}

The canonical partition function on the phase space in equilibrium case can writhe in the general form:
\begin{equation}
\rho(q,p)d\Gamma=\exp\{\frac{F-H(q,p)}{\Theta}\} d\Gamma
\end{equation}
where Hamiltonian of system $H(q,p)=E $ on the hyperspace constant energy $E$, $d\Gamma=\prod_{i}dq_{i}dp_{i}$ 
is element of the phase space $\Theta=kT$ where $T$ temperature and $F$ is free energy which can be determined from 
normalization condition $ \int \exp(\frac{F-H(q,p)}{\Theta})d\Gamma=1.$ As shown in \cite{Gibbs} the phase space depended 
only from energy and external parameters and if introduce the additional function $\Sigma=\ln\frac{d\Gamma}{d E} $ 
the canonical partition function can present in the other form:
\begin{equation}
\rho(E)dE=\exp\{\frac{F-E}{\Theta}+\Sigma(E)\}dE \label{11}
\end{equation}
which make in possible to describe the dependence of distribution function from energy of macroscopic systems. 
The normalization condition in this presentation can write in the form $ \int \exp(\frac{F-E}{\Theta}+\Sigma(E)) dE=1$
from which can determine the free energy of system taking into account the determinant transformation between phase space
and energy variable. In order to select the states  that give the dominant contribution to the partition function  we employ 
the condition $\frac{d\Sigma}{dE}=\frac{1}{\Theta}$ that determines the temperature of the system provided the change of the 
phase space as a function of the system energy is known. Using this definition and taking into account the basic principles 
of statistical mechanics \cite{Lan} we come to the conclusion that $\Sigma=\ln\frac{d\Gamma}{d E}= S $ is equal to the entropy 
of the system.  Now we have to make a very important notice: the temperature describes the dependence of the entropy only on 
energy, but not on any other thermodynamic quantities. We can define the temperature for other situations, but this definition 
has no sense without changing entropy.  Another important conclusion is that we can calculate the partition function by integration 
over energy. Such integration in this sense means the continual integral on  the energy variable . The extremum of the partition function 
is realized  under the condition: $F=E-\theta S$ and  any probable deviation from this condition gives very small contribution 
to macroscopic characteristics similar to the quantum contribution to classical trajectories  \cite{Lev1}, \cite{Lev2}. 

After that well-founded presentation of canonical partition function we can present recently proposed approach to description 
non-equilibrium macroscopic systems \cite{Lev1},\cite{Lev2}. There may be considered a natural hypothesis, that the 
non-equilibrium distribution function will depend on the energy of the system \cite{Gibbs}. In a similar way, our considerations will 
be based on the energy representation, where the states of the system are determined solely by their energies. The variation 
of system energy can induce the variation of the state of the macroscopic system. In the absence of any other knowledge about the 
non-equilibrium system, there is no reason to favor any state of system determined through energy as slowly parameter. The system energy 
variation define the state of the system. The non-equilibrium distribution function, as in equilibrium case, can be define as $\rho(E,t)$, 
which include the dependence on energy of the system $E$ and internal parameter ``time'', which indicate himself process. The energy 
distribution function, in general case, can be obtained from the basis kinetic equation, which present the system evolution during a 
long period of time and take into account the possible fast process in it. In general case, the non-equilibrium system can transit from 
one energy value to another one and this process depends on the external effect and initial conditions. The external effect present first 
of all in the variation of system energy, the energy also changes if the system dissipates or absorbs the energy as a result of the initial 
peculiarity of the behavior. The main idea of this presentation consists in the description of the evolution of a non - equilibrium system 
as a possible Brownian motion of the system between different states with changing energy and diffusion in the energy space \cite{Lev1}. 

In general case one can suppose, that dissipation ob absorption energy process can present at equation in standard form 
\begin{equation}
\frac{dE}{dt}=f(E)+g(E)L(t)
\end{equation}
This dissipation equation depend on external influence and initial conditions. The external influence, first of all,
manifests in change of energy of the system which dissipate or absorb by external action. This process is taken into account by
the first part of present equation which describe direct influence of the environment on the single macroscopic system.
This part can be obtain from dynamic of any macroscopic system if fully determine the directly interaction of this system with
environment. But this is not true in all case. However, in general case, the system exist in a contact with a nonlinear
environment. The system energy changed as result of random influence of the environment and this phenomena is taken into
account by the second part of the equation. The influence of this random influence is not correlated and the correlation 
between two values of fluctuation at two different moments $\left\langle L(t)L(t')\right\rangle=\phi(t-t')$ can be not zero 
only for time interval, which is equal to the time of action. The symbol $\left\langle ...\right\rangle$ mean the statistical 
averaging of appropriate value. The function $\phi \delta(t-t')$ must have the drastic peak at environs of zero and satisfy the 
condition $\int \phi(\tau)d\tau=\sigma^{2}$ for the white noise \cite{Kam} - \cite{Hor}. The system which can not rich the 
equilibrium after the fast changes of the environment must relaxing to the new state. This process indicate the possible 
degradation of the system in contact with nonlinear environment. 

The true nonlinear Langevine equation must have equivalent equation for probability distribution function which can be wrote,
according to physical process. At this time has been proposed two different approaches. If one consider that the coefficient
$g(E)$ depend on energy at start point, the equation for non-equilibrium distribution function can be obtained in the
Ito form. Also, if this coefficient depend on energy before and after transition, the diffusive equation can be written in the
Stratonovich form:
\begin{equation}
\frac{\partial \rho}{\partial t}=-\frac{\partial}{\partial
E}\left(f(E)\rho\right)+\frac{\sigma^{2}}{2}\frac{\partial}{\partial
E}g(E)\frac{\partial}{\partial
E}g(E)\rho
\end{equation}
Later on this article only Stratonovich presentation was used, because both presentation are connected \cite{Kam}-\cite{Hor}.
In described case, the different states of any system are determined and can be formed by previous state and possible
future states. The present equation for non-equilibrium distribution function in such case can be re-expressed in more
usual form of local conservation law for probability:
\begin{equation}
\frac{\partial \rho(E,t)}{\partial t}=\frac{\partial
J(\rho(E,t))}{\partial E}
\end{equation}
where the flow of probability can be written as:
\begin{equation}
J=-\left(f-\frac{\sigma^{2}}{2}g\frac{\partial}{\partial E}g\right)\rho+\frac{\sigma^{2}}{2}g^{2}\frac{\partial}{\partial E}\rho
\end{equation}
with diffusion coefficient $D(E)= \frac{\sigma^{2}}{2}g^{2}(E)$. The distribution function as stationary solution in 
the general non-equilibrium case can be presented as:
\begin{equation}
\rho_{s}(E)=A exp\left\{-U(E)\right\},
\end{equation}
with
\begin{equation}
U(E)=\ln \frac{g(E)}{g(E_{0})}-\int^{E}_{E_{0}}\frac{2
f(E')dE'}{\sigma^{2}g^{2}(E')}
\end{equation}
The $E_{0}$ indicate the initial state of system. This distribution function have the extremal value by energy, which can find as a 
solution of equation 
\begin{equation}
U'(\widetilde{E})=\frac{1}{D(E)}\left(D'(E)-f(E)\right)
\end{equation}
where $'$ stands for the energy derivative. This equation is equal to $D'(\widetilde{E})=f(\widetilde{E})$
which determine the relation between the system dissipation and diffusion in stationary case and completely determine new
stationary state of system. When the system dissipation $f(E)$ is represented by the nonlinear function of the state, and 
diffusion coefficient depend on the energy, a lot of interesting situations, including the noise-induced transition in new, 
more stable, stationary states, can be obtained. 

Let return to Universe formation. Can assume that initial vacuum state have the energy $E_{0}$ which produce fluctuations all 
possible field which can be born in such vacuum. First from all can determine the equation of state for vacuum. From thermodynamic 
relation can determine the pressure of vacuum as $P=-\frac{d E}{dV}$ where $V$ is volume for constant entropy. The entropy of 
initial state are fully determined an can take zero or other constant. From previous this thermodynamic relation
\begin{equation}
P=-\frac{d E_{0}}{dV}=-\rho_{v}
\end{equation}
where was assume that the additive energy $E_{0}=\rho_{v}V$ with density of energy $\rho_{v}$. It is general
equation of state for vacuum in all case and this relation determined only trough statistical principles if system
can change the possible states which induced the law of increasing of the entropy. This is general motive dynamic
of all states for macroscopic systems. When the system dissipation $f(E)$ is represented by the nonlinear function of the state, and diffusion 
coefficient depend on the energy, including the noise-induced transition in new, more stable, stationary states, can applied 
to description the Universe formation. 

\section{Universe formation}

First from all can assume that initial vacuum state have the energy $E_{0}$ which produce fluctuations all 
possible field which can be born in such vacuum. First from all can determine the equation of state for vacuum. From thermodynamic 
relation can determine the pressure of vacuum as $P=-\frac{d E}{dV}$ where $V$ is volume for constant entropy. The entropy of 
initial state are fully determined an can take zero or other constant. From previous this thermodynamic relation
\begin{equation}
P=-\frac{d E_{0}}{dV}=-\rho_{v}
\end{equation}
where was assume that the additive energy $E_{0}=\rho_{v}V$ with density of energy $\rho_{v}$. It is general
equation of state for vacuum in all case and this relation determined only trough statistical principles if system
can change the possible states which induced the law of increasing of the entropy. This is general motive dynamic
of all states for macroscopic systems. 

The Universe, as note previously are non-equilibrium \textsl{a priory}. If use the additional internal parameter as ``time'' 
we can obtain the very important relation as:
\begin{equation}
\frac{dS}{dt}=\frac{1}{\Theta}\frac{dE}{dt}
\end{equation}
from which assume that the changing entropy in time determine the changing energy from introducing internal parameter. For
real macroscopic system exist law of increasing of entropy and from thermodynamic relation $\frac{dS}{dt}>0$ and condition
that the phase space increase from changing of energy leads that $\frac{dE}{dt}> 0$ if system relax to equilibrium state 
as state with greatest entropy. If known the changing energy of system we can determine the internal parameter if 
are determined the entropy of system. The introduced internal parameter ``time'' determine the increasing of entropy 
for changing of the energy of the system. In general case the entropy of system only increase if energy increase. In 
thermodynamics, heat is energy that flows between degrees of freedom that are not macroscopically observable. If start
the changing of possible states of vacuum we can determine the heat changing as $\frac{dQ}{dt}=\frac{dE}{dt}$
from which assume that the changing heat determine the changing energy from introducing internal parameter and heating
will be possible only the changing of energy of system and system relax to equilibrium state. 

Exist possibility decrease the energy of vacuum state. Birth new field in starting vacuum state change energy of previous state
and came to new state the vacuum. After came into being the new field the energy of vacuum decrease and part this energy can go 
to stabilization of states of new fields if are non-linearity in behavior of this field. Can propose the next step existence such 
possibility. Standard cosmological models involve a scenario of Universe nucleation and expansion based on a scalar field which 
is of fundamental importance for the unified theories of weak, strong, and electromagnetic interactions with 
spontaneous symmetry breaking. What is possible obtain necessary behavior of fundamental scalar field. For such 
behavior can propose the next second paradigm - the every macroscopic states of systems go to smallest energy. 
In this sense the state of vacuum together with formed distribution of introduced new field can have smallest
energy as initial state of clear vacuum.  Let in this vacuum arise the scalar field which can increase this 
energy and we can write new energy in the form
\begin{equation}
E=E_{0}-\frac{\mu ^{2}}{2}\varphi ^{2}
\end{equation}
where second term take into account the symmetry energy from scalar field. The coefficient $\mu ^{2}$ describe 
the coupling of ground state with arise the new field. This coefficient can not be constant if take into account 
all possible fluctuation other field which can be appearance in vacuum. We can take into account this fluctuations 
next way. The partition function for this system can be write in the standard form
\begin{equation}
Z\sim \int D \varphi \int D\xi \exp \frac{1}{\Theta}\left\{ -E_{0}+\frac 12\mu ^2\varphi ^2+\xi \varphi^2 -\frac{\xi ^2}{2\sigma ^2}\right\}
\end{equation}
where we present coefficient $\mu ^2=\mu ^2+\xi $ as sum average value and fluctuation part which stochastic
change this value in interaction process scalar field with possible fluctuations other nature. Last part in
exponent present the energy the fluctuation another nature with disperse $\sigma ^2$. After continuum 
integration over all possible fluctuation $\xi$ and use the property of Gauss integral can obtain that the 
partition function can rewrite in the form 
\begin{equation}
Z\sim \int D \varphi \int D\xi \exp \frac{1}{\Theta}\left\{ -E_{0}+\frac 12\mu ^2\varphi ^2+ -\frac{\sigma ^2 \varphi ^4}{4}\right\}
\end{equation}
that satisfy the new energy $E_{0}-\frac 12\mu ^2\varphi ^2+\frac{\sigma ^2 \varphi ^4}{4}$ where
$V(\varphi)=-\frac 12\mu ^2\varphi ^2+\frac{\sigma ^2 \varphi ^4}{4}$ present well-known form
standard energy fundamental scalar field. The full energy new ground state and fundamental scalar field
can present as 
\begin{equation}
E=E_{0}-\frac{\mu ^4}{4 \sigma^{2}}+\frac{\sigma^2}{4}(\varphi^2-\frac{\mu ^2}{\sigma^{2}})^{2}
\end{equation}
which have the next asymptotic-if $\varphi=0$ we have the ground state energy $E_{0}$, if $\varphi^2=\frac{\mu ^2}{\sigma^{2}}$
the energy take the value $E_{0}-\frac{\mu ^4}{4 \sigma^{2}}$. This energy can be go to zero if $\sigma^{2}$ -dispersion
of fluctuations of field other nature go to zero. If $\sigma^{2}$ go to infinity energy of new state go to starting
energy of ground state. As conclusion we can affirm that in this understanding we have standard form the necessary
potential of fundamental scalar field but have other behavior the energy of vacuum at presence the scalar field. The 
coefficient of non-linearity in potential energy was formed only coupling the fundamental scalar field with fluctuation
the field another nature. After this comments can describe the stochastic scenario formation of Universe.

\section{Saddle state of Universe}

If return to description of Universe formation in the form Fokker-Planck presentation we must obtain the equation which 
describe the degradation of energy. In general case we can write $\frac{dE}{dt}=\frac{dE}{d \varphi}\frac{d\varphi}{d t}$
The dynamic equation for scalar field we can write in the standard form \cite{Lan} : 
\begin{equation}
\frac{d\varphi}{d t}=-\gamma \frac{dF(\varphi)}{d \varphi}
\end{equation}
where $\gamma $ is friction coefficient for dissipative fundamental scalar field. From theory of phase transition for fundamental 
scalar field  free energy $ F(\varphi \Theta)= V(\varphi,0) $ for zero temperature and previously equation transform to 
\begin{equation}
\frac{d\varphi}{d t}=-\gamma \frac{dV(\varphi)}{d \varphi}
\end{equation}
which are well known in standard cosmology \cite{LIN}.  If take into account that $\frac{dE}{d \varphi}=\frac{dV(\varphi)}{d \varphi}$ 
we can obtain the general Langevine equation 
\begin{equation}
\frac{dE}{dt}=-\gamma (\frac{dV(\varphi)}{d \varphi})^{2}=f(E) 
\end{equation}
with $f(E)=-\gamma(\frac{dV(\varphi)}{d \varphi})^{2} $. In the simple case when $V(\varphi)= -\frac 12\mu ^2\varphi ^2$ we get that 
$f(E)=- 2\gamma \mu ^2 E$. If take into account as previously the possible fluctuation of parameter $\mu ^2=\mu ^2+\xi $ we can 
determine $g(E)=2\gamma E$ and obtain the stationary solution in the form \cite{Hor}:
\begin{equation}
\rho_{s}(E,t)=N (\frac{E}{E_{0}})^{-(\frac{\mu ^2}{\gamma\sigma^{2}}+1)}
\end{equation}
This stationary solution is not equal to the solution in standard case, when the diffusion and friction coefficient do not depend on energy. 
If take another relation between diffusion coefficient and energy degradation $D'(E)=f(E)$ can obtain that $D=\frac{\gamma \mu^{2}\widetilde{E^{2}}}{2} $.
From Fokker-Planck equation we can obtain that $D=\frac{\sigma^{2}\widetilde{E^{2}}}{2} $ and assume that the friction coefficient 
depended from the intensity of fluctuation of the energy in system. This two relation can interpreted as relation between friction 
coefficient and intensity of fluctuations of the energy for simple model of fundamental scalar field. In more specific model this
relation can not take place. Without any calculation we can conclude, that preferable state for Universe are state with existence
scalar order parameter and with possibility formation the bubble of new state, which have the energy $E=E_{0}-\frac{\mu ^4}{4 \sigma^{2}}$. 
The entropy of ground state with energy of vacuum $E_{0}$ have the minimum entropy and increase with formation of bubble new phase.

It is possible to present the situation when the system energy increasing but exist mechanism which limit the energy. For the process 
which can be described in the term of dissipation function $f(E)=k^2 E-E^{2}$ second part take into account such limiting. The 
absorption parameter can be present as $k^2(t)=k^2+\xi_{t}$ where second part describe the random change of influence of 
environment. The Fokker-Planck equation take the form \cite{Hor}:
\begin{equation}
\frac{\partial \rho(E,t)}{\partial
t}=\frac{\partial}{\partial E}\left((k^2 E-E^{2})
\rho(E,t)\right)+\frac{\sigma^{2}}{2}\frac{\partial^{2}}{\partial
E^{2} }E^{2}\rho(E,t)
\end{equation}
The stationary solution of this equation can be wrote as
\begin{equation}
\rho_{s}(E,t)=N E^{-(\frac{2k^2}{\sigma^{2}}+1)}\exp\left\{-\frac{2}{\sigma^{2}}E\right\}
\end{equation}
which look like but not equal the thermal distribution function.

\section{Conclusion}

All cosmological model start from Big Bang and born the heat Universe. The presented approach can  change the starting
position. First from all exist only vacuum states with different fluctuation different field other nature. In the starting
``time'' when the born additional scalar field the number of possible states increase and energy of vacuum degrading and
formed bubble new phase grown. In this period work standard approach in cosmological model. For critical temperature \cite{LIN}
$\Theta_{c}=\frac{2 \mu }{\sigma}$ symmetry in behavior of fundamental scalar field restore and we have usual well-known
behavior of Universe, but with small difference. The initial vacuum not restore fully and we observe increasing acceleration
expansion of Universe. As show previously temperature symmetry restore depended from disperse of fluctuation of field other
nature but in the case formation bubbles new phase not all possible fluctuation can influence on the process inside the
Universe. This effect known as effect condensation of fluctuation \cite{Lev3}. The acceleration not stop never.
As show before when taking into account all possible multiplicative fluctuations and 
calculating the probability of transition into a stable vacuum state can be realized 
different situation. For this reasons we can employ the interaction with multiplicative 
noise which consider the possible changing as disperse of the fluctuations as the potential 
of the scalar field. In this case non-linearity must be taken into account in potential 
as in behavior of the fluctuations of other nature. Possible fluctuations change the minimum 
of potential and determine the other way of the dynamic Universe formation. For this behavior 
the state of the Universe in present case can determine $\varphi=0$ but non zero Hubble constant. 
Thus, has been proposed a model for describing the non-equilibrium Universe with determination 
of new stationary states. The stationary distribution functions of the Universe have been 
obtained for standard mechanism of formation, in contact with a nonlinear environment. That 
statement is well founded as presence of the negative mass coefficient in potential of the 
fundamental scalar field. The presence of the negative mass coefficient produce the non stability 
of ground state and the appearance of the possible fluctuations, which can be taken into account 
only in this approach.

\section{Acknowledgment}
This article carry out for financial support from theme department of physics and astronomy of NAS Ukraine 
``Dynamic formation spatial uniform structures in many-body system '' PK 0118U003535. This work was supported 
in part by Brain Pool programm by Grant N 218H1D3A2065894 through the National Research Foundation of Korea (NRF).

\end{document}